# Unprecedented decarbonization of China's power system in the post-COVID era

Biqing Zhu[1], Rui Guo[1], Zhu Deng[1], Wenli Zhao[2], Piyu Ke[1], Xinyu Dou[1], Steven J. Davis[3], Philippe Ciais[4], Pierre Gentine[2], Zhu Liu[1*]

1 Department of Earth System Science, Tsinghua University, Beijing 100084, China.

2 Department of Earth & Environment Sciences, Columbia University

3 Department of Earth System Science, University of California, Irvine, 3232 Croul Hall, Irvine, CA 92697-3100, USA

4 Laboratoire des Sciences du Climat et de l'Environnement LSCE, Orme de Merisiers 91191 Gif-sur-Yvette, France

*Corresponding Author: zhuliu@tsinghua.edu.cn

**In October of 2020, China announced that it aims to start reducing its carbon dioxide ($CO_2$) emissions before 2030 and achieve carbon neutrality before 2060[1,2]. The surprise announcement came in the midst of the COVID-19 pandemic which caused a transient drop in China's emissions in the first half of 2020[3]. Here, we show an unprecedented decarbonization of China's power system in late 2020: although China's power related carbon emissions were 0.5% higher in 2020 than 2019, the majority (92.9%) of the increased power demand was met by increases in low-carbon (renewables and nuclear) generation (+9.3%), as compared to only +0.4% for fossil fuels. China's low-carbon generation in the country grew in the second half of 2020, supplying a record high of 36.7% (+1.9% compared to 2019) of total electricity in 2020, when the fossil production dropped to a historical low of 63.3%. Combined, the carbon intensity of China's power sector decreased to an historical low of 519.9 $tCO_2$-eq$GWh^{-1}$ in 2020. If the fast decarbonization and slowed down power demand growth from 2019 to 2020 were to continue, by 2030, over half (50.8%) of China's power demand could be provided by low carbon sources. Our results thus reveal that China made progress towards its carbon neutrality target during the pandemic, and suggest the potential for substantial further decarbonization in the next few years if the latest trends persist. *[228 words]***

China's power system is a major source of the country's energy-related carbon emissions[3-6], accounting for roughly half of it's overall emissions, and 14% global energy-related emissions[4]. Lockdown measures applied during the 2020 COVID-19 pandemic (for example, promoting remote working and studying, closing down non-essential business) have dramatically changed power demand dynamics and reduced power related emissions[3,7]. Restrictions of in-person and commercial activities with reduced supply and consumers demand caused a 5.4% decrease of Chinese carbon dioxide ($CO_2$) emissions in the first half of 2020[3]. These impacts were however short-living, as rebound in electricity demand and $CO_2$ emissions, especially from the industry sector[3] was observed with the relaxation of restrictions.

In October 2020, for the first time, China announced its goal to reduce emissions before 2030 and to pursue a follow-up effort to achieve carbon neutrality by 2060, as strong efforts to meet the climate targets of the Paris Agreement, which calls for rapid and drastic reductions of global Greenhouse gas (GHG) emissions. Yet China's electricity system still relies heavily on fossil fuels (e.g., coal supplied 75% of all electricity in 2019), and substantial emissions reductions in this sector will thus need to overcome the lock-in of fossil infrastructure and scale-up deployment of low-carbon energy technologies[8,9]. Specifically, renewables (including hydro-electricity, solar power, wind power, geothermal power, biomass and other renewable sources) and nuclear energy will need to take up the share of coal in the energy mix as a long-term plan, to meet the energy demand while phasing out coal consumption. Several studies have analysed the decarbonization potential in China's power system, as to which extend and under which scenarios low carbon energy sources could replace the current share of fossil fuels as energy sources[6,10,11]. Scenarios based on historical developments prior to 2020 show that China needed to promote clean energy and to control coal-fired power, in order to peak the emissions earlier[12].

However, the COVID-19 pandemic has introduced new uncertainties and opportunities. The dramatically declined in power related emissions were closely linked to reduction in fossil-fired power[13]. The ongoing economic recovery and decisions about stimulus related to the pandemic will have a powerful influence on emissions due to short-term changes in consumption but also in the longer-term if these decisions are accompanied by changes in long-lived energy infrastructure.

According to China's Electricity Council, China's total power demand growth was above 3% in 2020 as compared to 2019 (https://cec.org.cn). This indicates that the temporal reduced power demand has quickly recovered. The COVID-19 pandemic has caused disruptive changes in everyday life for extended time period, but only led to limited reduction in carbon emissions, so far the long-term impacts on energy structure change (especially for renewables and nuclear energy) during and after the pandemic, and the driving factors behind changes have not been systematically evaluated.

Here, we constructed China's near-real-time carbon emission dataset with daily temporal resolution, covering source-specific power generation and sector-specific power consumption and their related emissions (See *Methods*). On the base of this newly constricted near-real-time power generation and related emission profiles from fossil (coal, natural-gas and oil /petroleum) and low carbon energy sources, we evaluate the short- and long-term impact of COVID-19 on the energy structure and power related emissions (nuclear power, hydro power, solar power, wind power, geothermal, biomass burning, waste burning and other renewables). We assess the climate impacts of changes in the electricity system by assess life-cycle GHG emissions with source-specific emission factors (see *Methods*). These emission factors account for direct fossil fuel $CO_2$ emissions and indirect emissions from construction, operation, land-use change, upstream and biogenic $CH_4$ emissions[14,15]. The driving factors behind power related emissions are evaluated by decomposing changes in energy structure and electricity demand from each of four main economic sectors (primary i.e. agriculture, secondary i.e. industry, tertiary i.e. service, and residential sector), which can be further disaggregated into 12 sub-categories. To separate the temporary impacts of climatic factors from other activity related impacts (such as the COVID-19 and infrastructure expanding), temperature and resource availabilities are analysed using satellite re-analysis data. In the end, we synthesis the abovementioned information to evaluate the long-term impact on China's power system and its associated emissions in the post-COVID era with scenario analysis. Details of our analytic approach, accounting methods and assumptions are described in *Methods* with the methodology developed previously[13,16].

**Dynamics in China's power generation and greenhouse gas emissions in 2020**

We found that although the year-end data show that generation increased by 253.3 TWh (+3.5%) in 2020, such rebound is with significant decreases in carbon intensity and fossil power share (Fig. 1). Figure 1b shows daily and cumulative changes in emissions. In total, the power sector produced 419.4 $MtCO_2$-eq GHG emission in 2020, an 18.64 $MtCO_2$-eq increase (+0.5%) from 2019. The increased demand for electricity in 2020 was mainly met by a surprising increase in low-carbon (non-fossil) generation, which increased by 235.0 TWh (+9.3%) from 2019—thus accounting for 92.8% of the increase in demand (Fig. 1a). In comparison, fossil power increased by just 18.3 TWh (+0.4%). The combined effect on the carbon intensity of China's power sector was a decrease to an historical low of 519.9 $tCO_2GWh^{-1}$ (averaged over 2020), along with the lowest share of thermal production (63.3%) and highest share of low-carbon electricity (36.7%) in the country's history (Fig. 1d, historical development see SI Fig.1a, b).

As the country where COVID-19 cases first emerged[17], there was a sudden decline in electricity generation and related emissions (Figs. 1a, 1b, and 1c; Fig. 2). However, Low-carbon electricity declined by only 4.1 TWh (-88.1 $GWhd^{-1}$ or -0.9%, as compared to 3.9% pre-COVID growth in 2020), much less affected by COVID-related restrictions in contrast to fossil electricity (dropping by 109.4 TWh, -2.3 $TWhd^{-1}$ or -11.4%, as compared to -6.4% pre-COVID decline in 2020). As a result, power related emissions declined by 11.6% during the COVID-19 pandemic (-116.1 $MtCO_2$-eq) as compared to 2019.

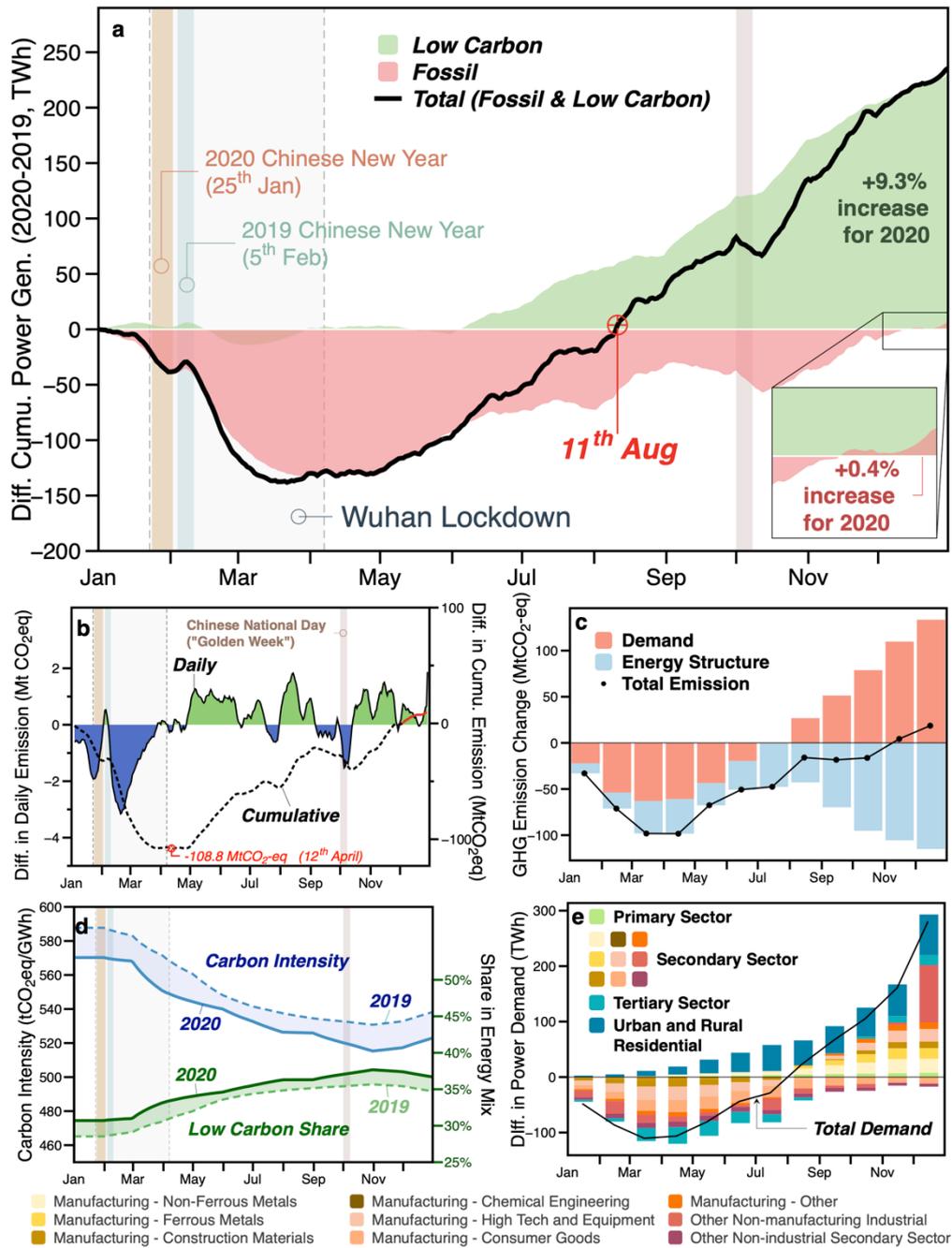

**Figure 1 | China's daily electricity generation and its associated carbon intensity. a**, Cumulative changes in electricity generation (black line), power from low-carbon sources (green) and from fossil source (red); **b**, change in daily and cumulative Greenhouse Gas (GHG) emissions; **c**, Decomposed changes in monthly emissions. Decomposition analysis (see *Methods*) of GHG emissions from China's power system. Demand refers to the power demand change. Energy structure refers to the change in each energy source sector's share in energy mix. Total emission refers to the difference between the change in total GHG emissions; **d**, cumulative carbon intensity (blue) and cumulative low carbon share (green) in the energy mix for year 2020 and 2019. **e**, Sector-specific power demand change. All data displayed in **a**, **b, c** and **e** show the change between 2020 and 2019. Wuhan lock-down refers to the movement's restrictions implemented due to the COVID-19 outbreak (23rd Jan - 7th Apr), and shaded in grey in figures. The Chinese New Year holidays of 2020 and of 2019 are indicated by orange and turquoise shades, respectively. The Chinese National Day (1st Oct) holiday ("Golden Week") is shaded in light purple.

Under the impact of COVID-19 (from January to March), decreased power demand account for 64% of the declined emission and improved energy (declined energy intensity, also see Fig. 1c) accounted for the rest 36% (Fig. 1b). Consumption reduction was observed for all sectors except for the urban and rural residential sectors (increased by 9.5 TWh or 3.4%). The biggest consumption decline is in the Secondary sector of economy (i.e. mainly industry) and the Tertiary sector (i.e. service), with 23.5 TWh (-8.2%) reduction in power demand (Fig. 1e).

Power demand rebounded steadily in the post-COVID time (since April, Fig 1.e), but its impact on emissions were offset by the continuously improving energy structure (shown by declined carbon intensity and increased share of low carbon power in energy mix in Fig. 1c). until the last days in December (Fig. 1c). By the end of 2020, Primary, Secondary, Tertiary sector and Urban and Rural Residential increased by 10.2%, 3.7%, 1.5% and 6.7% increase in power consumption, respectively. The only two sub-sectors of which the power demand did not recover to the 2019 level were Consumption goods manufacturing and other non-industrial secondary sector (declined by -2.3% and -5.1%, respectively, SI Fig. 4a). Contrary to the overall power demand increase (>1% for all main and sub-sectors), sector-wise carbon emission in the post-COVID recovery period stayed negative or remained comparable to 2019 for most sub-sectors, as a result of improved energy structure (SI Fig. 4b). Primary sector, Secondary sector and Urban and Rural Residential showed 7.0%, 0.9%, and 2.8% year-end increase, respectively, while Tertiary sector showed 1.4% year-end decrease in power related carbon emissions. In total, power associated emission increased by 4.1% in the post-COVID time as compared to 2019.

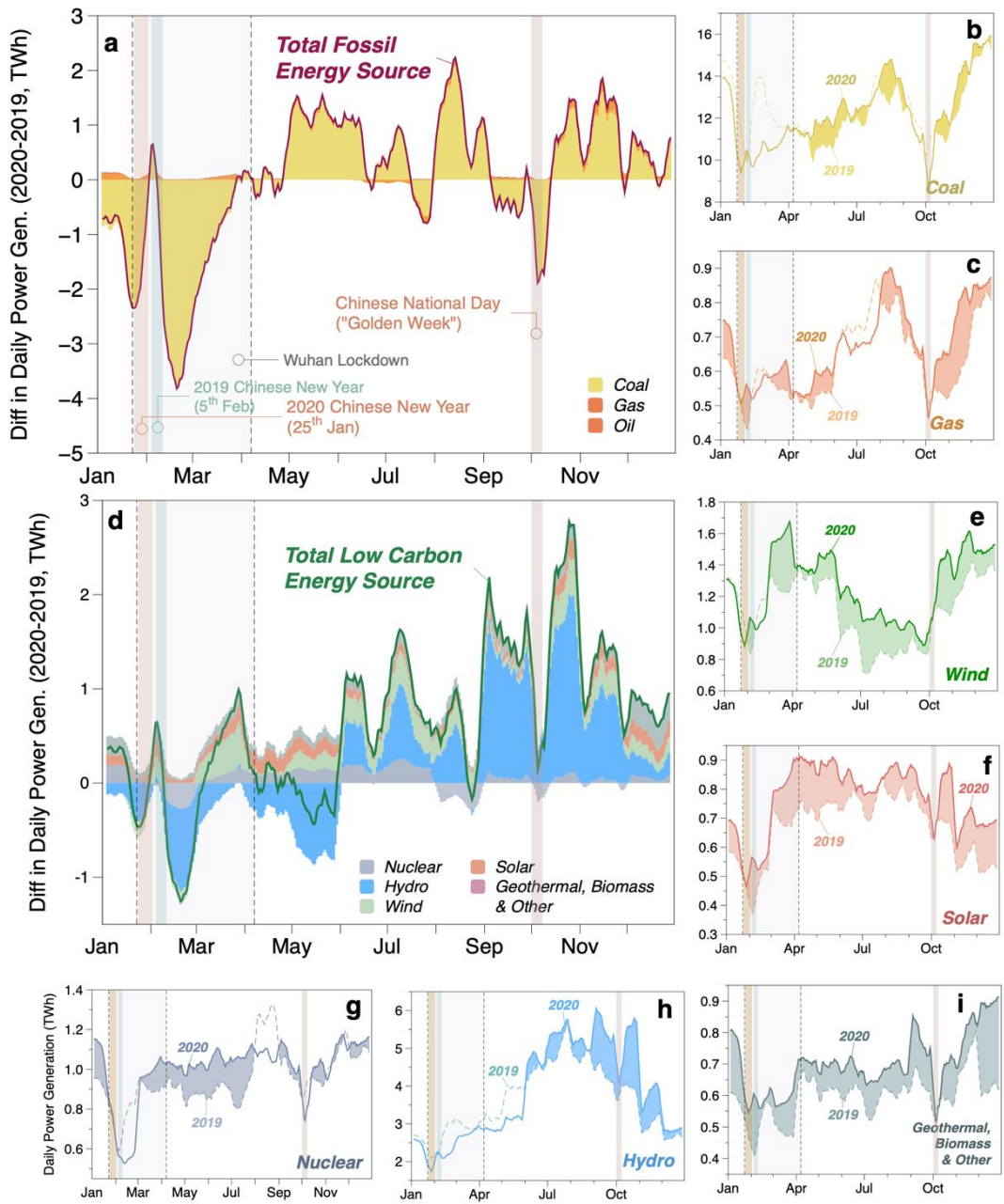

**Figure 2 | China's power generation from different or combined energy sources. a**, total fossil energy, **b**, coal, **c**, natural gas, **d**, total low carbon energy source, **e**, wind, **f**, solar, **g**, nuclear, **h**, hydro, and **i**, geothermal, biomass & other. Wuhan lock-down refers to the movement's restrictions implemented due to the COVID-19 outbreak (23rd Jan - 7th Apr), and shaded in grey in figures. The Chinese New Year holidays of 2020 and of 2019 are indicated by orange and turquoise shades, respectively. The Chinese National Day (1st Oct) holiday ("Golden Week") is shaded in light purple.

**Reduction of coal-fired power coincided with the pandemic**

More importantly, we show that COVID-19 lock-down further accelerated the reduction of coal from fossil power (Fig. 2 show daily variation with 2019 as baseline; SI Fig. 2 provide daily variation with 2016-2019 as baseline). Coal-fired power declined at an average rate of -1.4 TWhd$^{-1}$ during the lockdown (as compared to -1.1 TWhd$^{-1}$ decline prior to COVID-19 lock-down), to 15.4% below its 2016-2019 average and beyond its 2016-2019 variations (as compared to 0.6% above its 2016-2019 average prior to COVID-19 lockdown, SI Fig. 2). Nuclear power also showed decline coincided with the COVID-19 pandemic (by -19.7 GWhd$^{-1}$ and declined to 2.8% below its 2019 level but within its 2016-2019 variations, SI Fig. 2e). Hydro-electricity on average declined by -333.8 GWhd$^{-1}$ as compared to 2019 but stayed at the same level as 2016-2019 average (SI Fig. 2f) with no clear time correlation with the COVID-19 pandemic (Fig. 2h). Wind power, solar power, other renewables increased by 126.3 GWhd$^{-1}$, 103.6 GWhd$^{-1}$ and 86.2 GWhd$^{-1}$, respectively (Fig. 2e, 2f, and 2i, also above past five-year average and 95% confidence interval, see SI Fig. 2). In summary, the COVID-19 lock-down led to an immediate but short-term decline in coal and nuclear power, while gas-fired power, wind power, solar power and other renewables continued their increase.

In China's post-COVID recovery phase in 2020 (from the ease of Wuhan lockdown as of 8$^{th}$ April, till the end of 2020), low carbon power continued its steady increase with 847.5 GWhd-1 (11.6%), and accounted for 62.4% of the increased power demand (1.4 TWhd$^{-1}$, 6.7%, with -0.3% decline linked to temperature change). Power generation from every source showed net growth in post-COVID recovery, with coal power (448.3 GWhd$^{-1}$) and hydro-electricity (402.1 GWhd$^{-1}$) and the main contributor, followed by smaller but substantial growth in wind power (176.6 GWhd$^{-1}$) other renewables (106.3 GWhd$^{-1}$) and solar power (105.1 GWhd$^{-1}$). Nuclear power grew by 57.4 GWhd$^{-1}$, ranks as the smallest among low carbon power. The smallest growth is observed in gas (40.8 GWhd$^{-1}$) and oil (21.0 GWhd$^{-1}$). With the recovery, coal power resumed to the same level as the past four-year average, while all other power sources achieved growth that is above the past five-year average and 95% confidence intervals (SI Fig.2 c-i).

Combined, our calculation show there was rapid decarbonization and emission decrease in the power system during the COVID-19 lockdown, although the declined emission was largely offset by the demand increase in the post-COVID recovery, the decarbonization trend maintained as low carbon power grew steadily and strongly. The share of coal-fired power declined by 3.2% as compared to 2019 (also declined by 7.4% as compared to 2016-2019 average). Share of low carbon power increased by 2.5% to 34.3% as compared to 2019 (also increased by 6.7% as compared to 2016-2019 average). Emission declined by 12.2% as compared to 2019 (but increased by 6.3% as compared to 2016-2019 average). By the end of 2020, the share of coal-fired power still declined by 2.1% as compared to 2019 (also by 6.0% as compared to 2016-2019 average. Share of low carbon power increased by 1.9% as compared to 2019 (also by 5.7% as compared to 2016-2019 average).

**Scenarios of Emission from China's power system**

Total life-cycle GHG emissions from eight energy sources are considered, with emissions from 2010 to the present, and projections to 2030 under various power demand and generation scenarios (Fig. 3 and over all year-over-year power generation dynamics from January, 2010 to December, 2019 see SI Fig5). For every month of the year from 2010 to 2019, China's total power demand, and source-specific power generations have followed strict near-linear growths for the past decade, as guided by the growth goals outlined in the "Five Year Plans" (the twelfth and the thirteenth). Total power demand shows the highest linearity ($R^2 > 0.97$ for every month). Nuclear power, wind power and solar also show very strong linearity in its year-over-year generation growth ($R^2 > 0.9$ for every month), following by fossil-fired power (from coal, gas and oil) and by power from other renewables ($R^2$ between 0.84 ~0.95). The growth of hydro power generation shows strong linearity in spring, summer and winter time ($R^2$ between 0.88~0.95), but less strong for the three months in autumn ($R^2 = 0.81, 0.69, 0.84$, respectively).

Given the fact that the power generation have followed strict near-linear growths for the past decade under the "Five Year Plans"（Fig. 3）, we can make a straightforward comparison to show the impacts on China's power demand and generation from the pandemic. We compare the impacts of COVID-19 on China's power system's energy structure and its related emissions for the next 10 years (from 2020 to 2030), by defining three scenarios of different power demand and policy directions. These scenarios are: First, baseline (BS) scenario, which assumes the continuation of the most recent development plan (the recently completed "13th Five Year Plan" for year 2015-2020) in fossil and low carbon power development for year 2020-2030. The BS scenario is constructed with 2019 as baseline, also referred to as "Scenario without COVID-19". Second, improved energy structure (IES) scenario, assumes the power demand will continue as before 2020 but the above-usual year-over-year fast expanding of low carbon power in 2020 will continue for 2020-2030. Third, reduced demand with decarbonization (RD), which assumes both the constrains of the power demand growth and the year-over-year fast expanding of low carbon power in 2020 will continue for 2020-2030. The baseline year of both IES and RD scenarios is 2020. Key assumptions are summarized in Table 1.

We acknowledged the limits and uncertainties of the scenarios, as previous studies suggested different predictions on the peak of China's energy demand ranging from 2025, 2030, 2035 or beyond[10,18-20]. However, according to China's Electricity Council, there is still projection of power demand growth from 2020 – 2035. Thus, a near-term scenario comparison based on the most recent plan and status of China's demand and generation give us insights on the potential for substantial further decarbonization in the next few years if the latest trends persist. In addition, power demand outlook and efficiency improvements in the BS scenario are consistent with trends in previous 'market-based' predictions, which assumed that current policies continuate and future renewable costs decreases moderately[6,10]. The structure outlook under the BS scenario is also very similar to other predictions assuming a continuation of current policies and developments. For example, the BS scenario predicts that China's renewables will increase by 1596 TWh by 2030, while IEA predicts an increase of 1 500 TWh[21].

Under the impact of COVID-19, the year-over-year growth of China's 2020 total power demand was below its previous growth rates (Fig. 4). The demand reduction was disproportionally reflected in coal-fired power, while low carbon power remained its high growth rate, resulted in a decarbonized power system as compared to its own past. Though China's power system is still heavily relying on coal as major power source, the recent "14th Five Year Plan" outlines the importance of limiting coal-fired power as crucial pathway to achieve carbon neutrality by 2060[6].

Under all three scenarios, China's total power demand will continue to grow at different rates (SI Fig 5), as well as its total power related GHG emissions (Fig. 4). Under the BS scenario, China's power related GHG emissions are yet to growth by 30.3% in 2030 (as compared to 2020, Fig. 3a, b), with carbon intensity reduced to 474.9 tCO$_2$eq/GWh (as compared to 525.3 tCO$_2$eq/GWh in 2020 under BS scenario, and 519.9 tCO$_2$eq/GWh based on actual data from 2020, Fig. 3d). While RD scenario predicts a small growth in power related emission (4.7% as compared to 2020), carbon intensity will also reduce to 407.5 tCO$_2$eq/GWh (21.6% decline as compared to 2020, Fig. 3c). In comparison, the increase in total power generation are close for RD (33.6%) and for BS (42.6%) scenarios. This is because under the RD scenario, 92.9% of the increased power demand will be met by low carbon energy sources, while in the BS scenario only 56.2% of the increased demand is provided by low carbon energy sources.

Compared to BS and RD scenario, the improved energy structure (IES) scenario is a more realistic reflection of China's power system in post-COVID era. It accounts for the high growth potential as predicted from historical data (Fig3 and SI Fig 5), as well as the decarbonized power system under the impact of the COVID-19. The reduced power demand growth in 2020 is closely related to the COVID-19 pandemic, which is unlikely to continue as the economy recovers (Fig. 1). In contrast, the fast-increasing low carbon power generation is linked to speedy expansion of China's low carbon generation capacities (Fig. 5). As emphasised by the "14th Five Year Plan", this expansion will continue in the near future as part of China's national policy towards carbon neutrality in 2060. IES evaluated the potential GHG emission with continuous increasing power demand and fast increasing low carbon power (same speed as 2019-2020). Results show that even with the high power demand growing potential as 2010-2019 (towards $1.08 \times 10^4$ TWh, thus 42.6% in power demand by 2030), GHG emission will only increase by 18.9%. By 2030, nearly half of the power demand can be provided by low carbon energy sources (Fig. 3e).

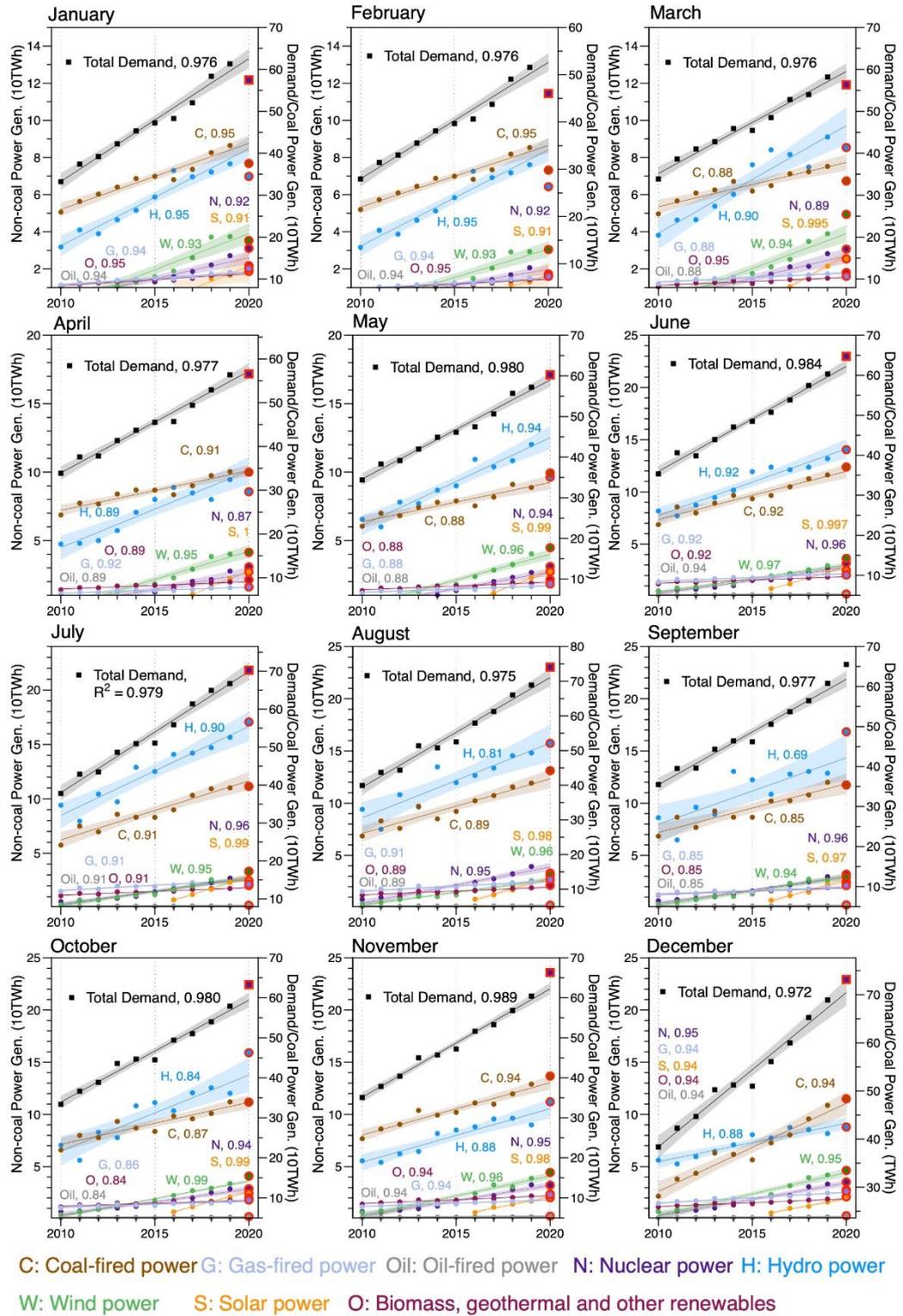

**Figure 3 | Source-specific year-over-year power generation for 12 months from 2010 to 2020.** Each panel represents one month. Every power source is represented by coloured dots. Total demand is represented by black dotes. Smoothed lines with the corresponding colours represent the fitted linear regression models of the source-specific power generation growth and year for each month. Shaded areas

represent the 95% confidence interval of the fitted lines. Linear regression models are fitted with data from 2010 to 2019. Data points in 2020 are highlighted red coloured circles. Numbers are the R-squared value, $R^2$

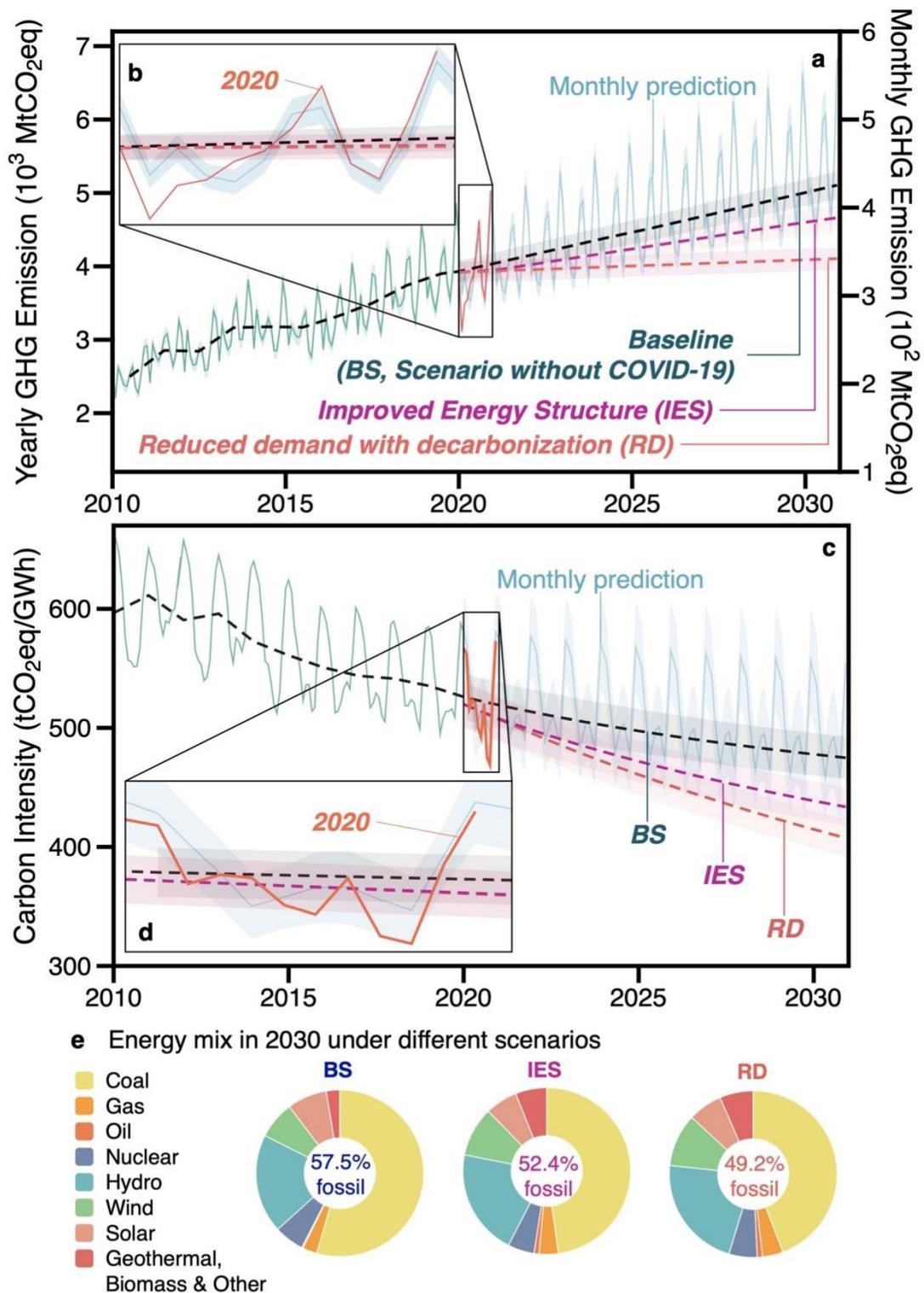

**Figure 4 | Scenario analysis on China's power related Greenhouse Gas (GHG) emissions, carbon intensity and energy mix. a**, China's power related GHG emissions from 2010 to 2030 under different scenarios (**b**, for 2020). Green line shows the historical carbon emission from 2010 - 2019. Blue line shows the baseline emissions without COVID-19. Shaded blue area shows the 95% prediction interval.

Dashed purple line indicates the carbon emission based on improved energy structure (IES) scenario. Dashed red line assumes the demand and source-specific power generation continue to change at the same magnitude as 2019-2020 (RD scenario). **c.** Carbon intensity of China's power sector from 2010 to 2030 under different scenarios (**d**, for 2020). China's power related GHG emissions in 2020. Solid red line shows China's actual power related emission. **e**, Energy mix in 2030 under different scenarios. From left to right, pie charts represent the energy mix of China's power system in 2030 under BS, IES and RD scenarios, respectively. The share of fossil-fired power in the power system are noted in coloured texts.

**Table 1 Model scenarios**

|  | **Baseline (BS)** | **Improved energy structure (IES)** | **Reduced demand with decarbonization (RD)** |
|---|---|---|---|
| Base year | 2019 | 2020 | 2020 |
| Future power demand assumptions | Continuation of 2010 – 2019 year-over-year growth rate | | Same growth rate as 2019 – 2020 |
| Non-coal power generation assumptions | Continuation of most recent development plans | Same growth rate as 2019 – 2020 | |
| Coal-fired power generation assumptions | Continuation of most recent development plans | Fill in the power deficiency between power demand and non-coal power generation | Same growth rate as 2019 – 2020 |

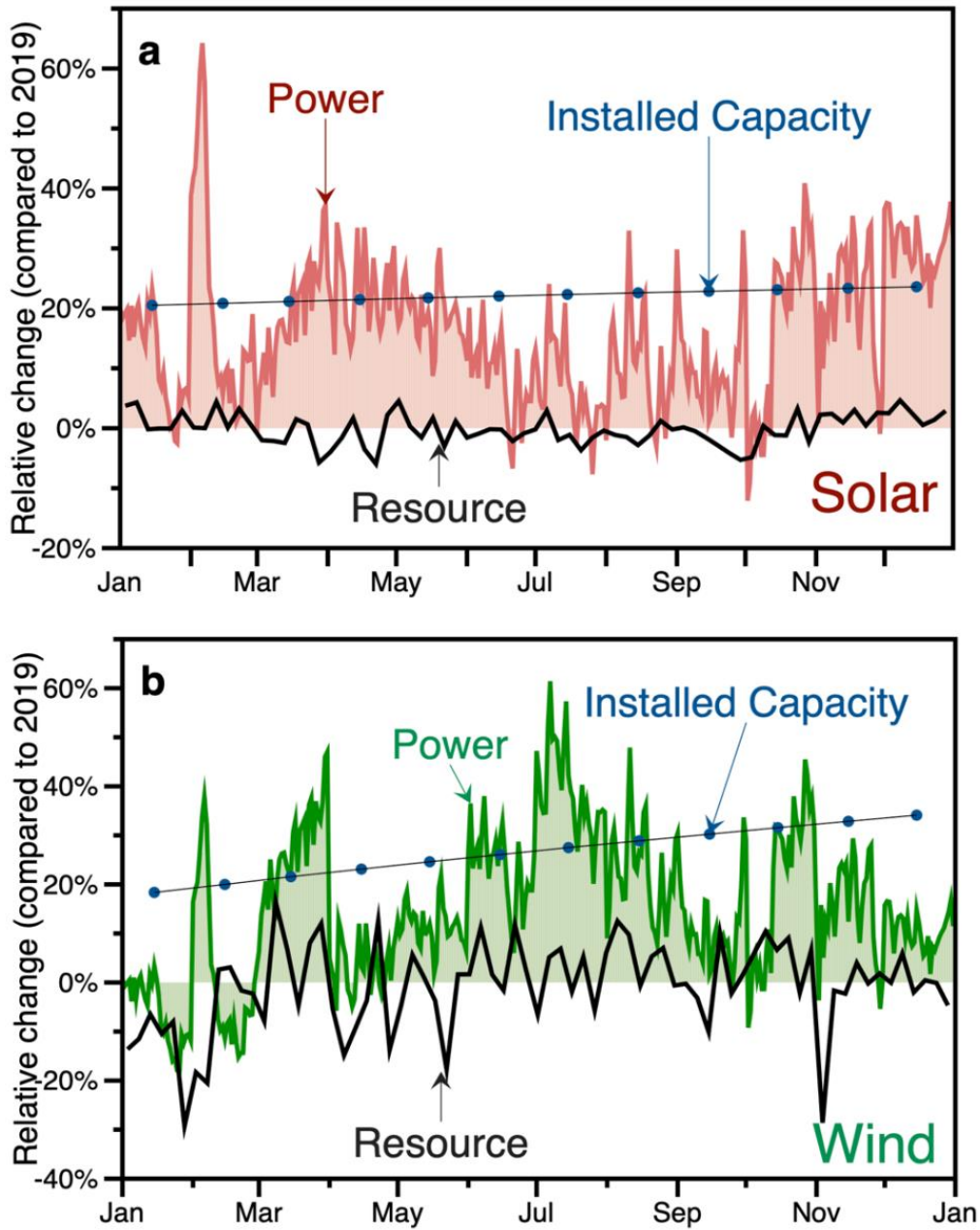

**Figure 5 | Resource availability, installed generation capacity for wind and solar power.** Bar graphs indicate daily power generation change. Black lines indicate resource availability change (calculated from near surface wind speed and surface solar irradiation). Blue dots indicate installed generation capacity change. All changes refer to year-over-year relative change, expressed in percentage. Data for January to October are averaged over 5-days interval. For November and December, interpolation and prediction methods were used to account for compromised data quality due to intensive cloud coverage.

**Climatic factors and uncertainties**

In order to discriminate the impact of COVID-19 pandemic from other factors such as temperature abnormality and resource availability. Power demand change can be partially explained by abnormally warmer or colder temperature. In general, there is medium to strong correlation between temperature and daily power generation (SI Fig. 3). With this correlation, we found that temperature could only explain a small fraction of power generation variation: among the 3.5% increase in year-end total demand, temperature accounted for -0.5% (details see *Methods*, temperature abnormality and correlation between temperature and power generation see SI Fig. 3, monthly temperature related power change see SI Table.1). Other climatic factors also played limited roles in the change. For example, wind and solar power increased by 14.1% and 15.9%, respectively, while wind and solar resources availability showed only 0.9% increase and 0.52% decline, respectively (analysed by corrected wind speed[8,22,23] and solar radiation[23] reanalysis data, details see *Methods*, relative change over time see Fig. 5). The increased wind and solar power production are in line with the fast-expanding power plants due to stimulating policies [9], which led to 34.6% and 24.1% increase in wind and solar installed generation capacity, respectively (Fig. 5).

As there are no other database provides up-to-date daily power generation and emission data for China, it is therefore difficult to evaluate the uncertainty of our constructed near-real-time power generation and emission data based on cross-compression methods. Nevertheless, we performed uncertainty analysis with available monthly or yearly data from literature (detail see *Methods* and ref[3]). The overall estimation of total power generation show small uncertainty (± 1.1 %). However, it is ± 4.4% for total emissions, mainly comes from uncertainties embedded in emission factors.

For scenario analysis, as we projected uncertainty derived from year-over-year variabilities and from emission factors. The abnormalities in power demand and generation from 2010-2019 are considered for future projections. Yet, there are still uncertainties in further projections. The power sector is a dynamic system, which is also affected by costs, storage and other process such as power transport. In order to achieve a constant or faster growth of low carbon power, resource capacity and transmission infrastructure must be scaled up quickly, to account for the imbalance between temporal and spatial imbalance between low carbon power resource and power demand. There is also uncertainty associated with emission estimation. Although the emission factor uncertainty is already considered in this study, the ranges of emission factors in this study are estimated based on current available technologies. It is possible that with technology advancement, there will be further decline in power related emissions and in carbon intensity.

## Discussion and Conclusions

The accelerated growth of low-carbon energy during the COVID-19 pandemic in 2020 show that low-carbon energy, due to its low marginal costs and low operational requirements, are the preferred option under such emergencies. The continuous growth of low-carbon energy in the recovery phase also further exhibit its robustness even with rebounded demand. There are still greater potentials for low-carbon energy technologies in China. Studies have shown that there is still great resource availability for China's low-carbon energy, such as hydro-electricity[24], wind[8], and solar power[25]. The upcoming "14th Five-Year-Plan" will entail further expansion of low-carbon energy infrastructure and greater shares of low-carbon energy sources in the energy mix, as key means for achieving China's carbon neutrality goal.

We show that with the same development rate as observed in the past decade, there will be substantial growth of power related emission from China. Urgent actions are needed to prevent this growth and to reduce carbon emission, including both to cap the power demand in the near future and to further expand low carbon power capacities. The marginal cost of renewables will decline further as more renewables installed. Wind power alone has the potential to meet projected increases in China's electricity demand between now and 2030, which could in turn reduce carbon emissions by 30% over the same period[8]. Studies suggested that 62% of China's electricity could come from non-fossil sources by 2030 at a cost that is 11% lower than if achieved through a business-as-usual approach. Further, China's power sector could cut half of its 2015 level carbon emissions at a cost about 6% lower compared with business-as-usual conditions[6]. To date, the speed of grid expansion has lagged behind new renewable capacity, so that China is deploying more renewable technologies than it can effectively use. For example, substantial quantities of wind generation are curtailed in rural China. Further investments in the national grid are thus urgently needed to facilitate continued growth in renewable generation. Coordination of energy storage may eventually be similarly important.

The COVID-19 pandemic will have a long-lasting effect on the electricity system. For example, although restrictions have been relaxed, residential electricity consumption has remained high, suggesting persistent behavioural changes of people staying home. Similarly, despite recovering at the end of 2020, demand in the tertiary sector (i.e. service sector) experienced a long lag period as compared with other sectors. The electricity system may eventually change accordingly. For example, decentralized generation may be well-suited to larger residential demands, but more localized grid management and power storage system will be required to balance the difference between power generation and demand peaks. As the world recovers from the pandemic and economic growth resumes, China's climate goals leave little margin for delay. The country must continue and accelerate both its deployment of low-carbon generation and its retirement of coal power plants. Despite modest growth in Chinese emissions in 2020, the trends and drivers of those emissions suggest that decarbonization in the country has begun; the next few years are critical.

# Methods

**Power data compiling, assimilation and simulation**

Source-specific electricity generation was calculated from source-specific monthly installed capacity multiplied by corresponding monthly full load hours. We compiled monthly installed capacity data and full load hour data from China Electricity Council when available

(https://cec.org.cn/detail/index.html?3-291651). When such monthly data is not available, we acquired supplementary monthly source-specific electricity generation data from the National Bureau of Statistics of China. China Electricity Council data was preferred over the other because it provides information on all in-grid electricity generation, while the other source only includes electricity generation from power plants over certain scales. For Solar electricity, for example, China Electricity Council provides information on both centralized and distributed photo-voltaic, while National Bureau of Statistics only covers power generation from centralized photo-voltaic. Exclusion of distributed photovoltaic, however, will lead to serious underestimation of total solar power generated in China, as the corresponding installed capacity accounts for nearly 50% of China's solar power system.

The power sources of which China Electricity Council provides detailed information on are thermal power, coal-fired power, gas-fired power, nuclear power, hydro-electricity, in-grid wind power, total photo-voltaic and distributed photo-voltaic (list and relationship with terms used in this study see Table 2). Thermal power includes power generated from coal, natural-gas, oil(petroleum), biomass, waste burning and power generation from residual pressure and residual temperature. After substituting coal and gas-fired power from thermal power, the rest of thermal power was dis-aggregated into oil-fired power and biomass, waste burning and power generated from residual pressure and residual temperature. Oil-fired power was further dis-aggregated from the other sources with additional monthly and yearly available fraction data from IEA[26] and BP[27].

**Table 2| Energy sources included in this study.**

| Terms used by CEC | | Energy source | Energy source described in this study |
|---|---|---|---|
| Thermal power | Coal-fired power | Coal | Coal |
| | Gas-fired power | Natural gas | Gas |
| | Coal, natural-gas, oil(petroleum) | Oil(petroleum), | Oil |
| | Biomass, waste burning and power generation from residual pressure and residual temperature | Biomass burning, waste, residual pressure and residual temperature | Biomass, Geothermal & other |
| Other | | Geothermal and other sources | |
| Nuclear power, | | Nuclear power | Nuclear |
| Hydro-electricity | | Hydro power | Hydro |
| Wind-electricity | | Wind power | Wind |
| Total photo-voltaic and distributed photo-voltaic | | Solar power | Solar |

We acquired daily thermal power mainly from State Grid, China from January 2018 till September, 2020. For months when direct daily thermal power data were not available (prior to 2018 and data gaps after 2018), we further estimate the daily thermal power production with the daily coal consumption of the Zhejiang Power Corporation (https://www.wind.com.cn/) with polynomial linear regression. Then this daily coal production data was used to disaggregate the monthly thermal power generation data from the National Bureau of Statistics (https://data.stats.gov.cn/). Following, daily source-specific power production data was generated with the daily-thermal power data and the pre-compiled monthly source-specific power production data.

**Life-cycle Emission and Carbon intensity calculation**

Lifecycle analysis is used to analyse the GHG footprint from the whole life cycle and its supply chain[28], including the upstream energy consumption and the fuel use for plant building, operation, and infrastructure etc[15]. Thus, in this study, we assess the life-cycle emissions from the power sector by multiplying the electricity production with the corresponding lifecycle emission factors (see Table 3) by different technologies (Eq. 1).

$$GHG\ Emissions\ = \Sigma(Power\ Generation_i \times Emission\ Factor_i) \qquad (1)$$

Here, the lifecycle GHG emission factors adopted in this study are collected from the IPCC report *Renewable Energy Sources and Climate Change Mitigation* in 2010 (ref [29]). The emission factors by different power generation technologies were reviewed and summarized by the National Renewable Energy Laboratory (NREL) based on near 300 screening publications of lifecycle assessments of electricity generation technologies. All lifecycle emission factors collected from the publications were categorized by technology and converted to the same unit.

It should be noted that the lifecycle emission factors exclude the land-use-change-related and heat-related emissions[30]. The lifecycle GHG emissions in this study only focus on the electricity production-related activities in its supply chain.

The carbon intensity of energy mix is calculated by the following equation (Eq. 2), defined as the GHG emitted per unit of electricity produced:

$$Carbon\ Intensity\ = GHG\ Emissions\ /\ Electricity\ Production \qquad (2)$$

**Table 3 | Life cycle emission factors by energy types. Data acquired, synthesised and adopted from IPCC special reports[29,30], which carried out a comprehensive review from lifecycle assessment publications of different electricity generation technologies. The range of hydro power emission factors are adjusted based on literature[31], with a focus on hydroelectricity facilities based on run of river and non-tropical reservoirs, excluding hydroelectricity facility types which are not commonly found in China, for example the ones in tropical regions.**

| Energy type | Life cycle emission factor (ktCO$_2$-eq/TWh) |
|---|---|
| Coal | 820 (ranging from 740~910) |
| Gas | 490 (ranging from 410~650) |
| Hydro | 24 (ranging from 0.5~152) |
| Nuclear | 12 (ranging from 3.7~110) |
| Solar (rooftop) | 41 (ranging from 26~60) |
| Solar (utility) | 48 (ranging from 18~180) |
| Geothermal | 38 (ranging from 6~79) |
| Wind (onshore) | 11 (ranging from 7~56) |
| Wind (offshore) | 12 (ranging from 8~35) |
| Biomass | 230 (ranging from 130~420) |

**Power demand data compiling, assimilation and associated emission calculation**

China power consumption data for 2019 and 2020 were derived from China Electricity Council electricity consumption and quarterly statistical analysis briefing reports (https://cec.org.cn/menu/index.html?303). The total consumption was categorized into those consumed in the primary sector, the secondary sector, the tertiary sector as well as the residential (urban and rural). According to the reports, we analyzed the electricity consumption distribution of the secondary sector in greater detail, mainly the industry sector and its major component the manufacturing industry. This is meaningful in that the industry sector is still the largest consumer of electricity in China[32], hence the changes of consumption structure and pattern within it will bear the greatest policy implication of electricity sector transform. Further, in order to gain more insights into the sectoral effects on electricity consumption and associated emission from the demand side, the manufacturing industry was divided into high-tech and equipment manufacturing, consumer goods manufacturing, other manufacturing and the so-called 'four energy-intensive manufacturing', the latter of which includes ferrous and non-ferrous metal, construction materials and chemical engineering sectors.

For May to November 2020, we directly obtained the consumption data and the year-on-year change rates from the cumulative monthly reports according to the divisions mentioned above. As such, the corresponding 2019 monthly consumption data were explicitly calculated as follows:

$$C_{(i,2019)} = C_{(i,2020)} * r_{i,y-o-y} \quad (i: May - Nov) \quad (3)$$

where $C$ is the consumption in 100 GWh and $r$ is the year-on-year (*y-o-y*) change rate. On the other hand, except for the 'four energy-intensive manufacturing' sector, the monthly data for the other three manufacturing sectors prior to May 2020 were not available publicly. Therefore, we assumed that for January to April 2020 the relative share of each of the three sub-sectors is the same as the average between May and November, in estimating the consumption data before then. Hence, we have:

$$C_{m,n} = C_{(m,manufacturing)} * \left(1 - \frac{C_{(m,four\ energy-intensive)}}{C_{(m,manufacturing)}}\right) * \frac{\sum_i C_{(i,n)}}{9} \quad (m: January - April; n: high-tech\ and\ equipment, consumer\ goods\ \&\ others) \quad (4)$$

To calculate the corresponding consumption data for 2019, we utilized the quarterly briefing data (https://cec.org.cn/detail/index.html?3-281670) for consumption from January to March and the year-on-year quarterly change rate, and further assumed that the distribution between January and February (see below) and March was the same as 2020. Moreover, because China's statistical reporting follows the convention that January and February data were always aggregated, we applied the generation data, which were separate for these two, as the proxy to dis-aggregate the combined data, assuming that the generation distribution between January

and February was preserved in electricity consumption as follows:

$$\frac{G_{Jan}}{G_{Feb}} = \frac{C_{Jan}}{C_{Feb}} \quad (5)$$

where G stands for the generation data.

In calculating the CO$_2$ emissions driven by the demand side, we firstly adopted the emission factors for each electricity resources i.e. coal, nuclear, solar etc. from the IPCC 2014 report[33] and then obtained the weighted carbon intensity (gCO$_2$eq/kWh) for each month between 2019 and 2020, based on the sectoral electricity generation data available as shown below:

$$CI_{(i\ \&\ m)} = \frac{\sum_x EF_x * G_{(i\ \&\ m,x)}}{\sum_x G_{(i\ \%\ m,x)}} \quad (x: coal, nuclear, solar\ etc.) \quad (6)$$

where *CI* and *EF* stand for carbon intensity and emission factor, respectively.

The monthly CO$_2$ emissions associated with total and sectoral consumption were then calculated by multiplying the carbon intensities by the corresponding consumption data (activity). As such, it was effectively assumed that the generation portfolios are applicable in each sector.

**Climate factors and temperature effect**

Wind and solar resource availabilities are estimated from wind speed and surface solar radiation.

The wind speed (WS) and solar radiation (Rs) are extracted from hourly ERA5 reanalysis dataset (The dataset can be downloaded from https://cds.climate.copernicus.eu/cdsapp#!/dataset/reanalysis-era5-single-levels?tab=overview). The spatial resolution is 0.25° x 0.25°. We extracted the China mainland area and aggerated the hourly data to 5 days average and monthly average to reduce noise and show the seasonal cycle of Rs and WS for year 2018, 2019, and 2020 （Fig. 5）.

The temperature data is acquired from ECMWF atmospheric reanalyses data[34]. The relationship between power demand and temperature change was expressed with linear regression models between daily fossil power generation and daily averaged temperature, after removing extended periods of holidays (Chinese New Year holiday and Chinese National Day holiday). This model was established and tested with data from 2018 and 2019. By taking the temperature difference between 2019 and 2020, the power demand change caused by temperature fluctuation was estimated with the established linear regression models.

**Index decomposition Analysis**

We adopted the Logarithmic Mean Divisa Index (LMDI) decomposition method[35] to perform the residual-free decomposition of the factors affecting GHG emissions related to China's

power generation. The adopted expression[35,36] is shown in Eq. 7:

$$E = \sum_i E_i = \sum_i Q \frac{Q_i}{Q} \frac{E_i}{Q_i} = \sum_i Q S_i I_i \tag{7}$$

Where $E$ is the total power related GHG emission (expressed as CO2-eq), Q is the gross power generation, and $S_i$ and $I_i$ are the shares of power source $i$ generation and emission factor (CO2-eq/generation) of power source $i$, respectively. The contribution of each driving factor causing the emission change (difference in GHG emissions between the $t^{th}$ month of 2020 and the corresponding month of 2019) can be expressed as the following equation (Eq. 8):

$$\Delta E_t = E^t - E^0 = \Delta E_{act} + \Delta E_{str} + \Delta E_{int} \tag{8}$$

Where $\Delta E$ is the difference between the GHG emissions from the $t^{th}$ month of 2020 ($E^t$) and the GHG emissions from corresponding month of 2019 ($E^0$). The contribution from activity, energy structure and emission intensity are noted as $\Delta E_{act}$, $\Delta E_{str}$, and $\Delta E_{int}$, respectively. Activity effect is indicated by power generation. Energy structure effect is indicated by the share of each power source in the energy mix. Emission intensity effect is indicated by the energy source's emission factor, which is shown in Eq.9 - Eq.12:

$$\Delta E_{act} = \sum_i w_i \ln\left(\frac{Q_i^t}{Q_i^0}\right) \tag{Eq.9}$$

$$\Delta E_{str} = \sum_i w_i \ln\left(\frac{S_i^t}{S_i^0}\right) \tag{Eq.10}$$

$$\Delta E_{str} = \sum_i w_i \ln\left(\frac{I_i^t}{I_i^0}\right) \tag{Eq.11}$$

$$w_i = \frac{E_i^t - E_i^0}{\ln E_i^t - \ln E_i^0} \tag{Eq. 12}$$

Where $Q^t$, $S^t$, and $I^t$ are the power generation, energy structure, and emission factor of the $t^{th}$ month, respectively. $Q^0$, $S^0$, and $I^0$ are the power generation, energy structure, and emission factor of the baseline month, respectively.

**Uncertainty Analysis**

We followed the 2006 IPCC Guidelines for National Greenhouse Gas Inventories[37] to conduct the uncertainty analysis for activity (power generation) and emission data. First, we estimated the uncertainties embedded in activity data considering the accuracy of our database, and by comparing with other widely used database[26,27]. As non-other database provides up-to-date information on daily source-specific power generation, our uncertainty estimation is based on monthly or yearly data when available. The uncertainty in activity data mainly comes from the difference between databases. The uncertainty of total-power (± 1.1%) is higher than the uncertainty of fossil-fired power (± 0.6%), but lower than the uncertainty of non-fossil low carbon sources (± 4.5%). The largest uncertainty comes from oil-fired power (± 24.4%) and biomass, geothermal and other renewables (± 27.7%). Coal-fired power, as the largest share in the power mix, has only 0.7% uncertainty.

The uncertainty in the emission is estimated by propagating the uncertainty from activity data and the uncertainty from emission factors. We combined uncertainty from all power sources and computed the uncertainty of total emission with the error propagation equation from the IPCC (Eq. 13).

$$U_{total} = \frac{\sqrt{\sum (U_s * \mu_s)}}{|\sum \mu_s|} \qquad (13)$$

Where $U_s$ is the share of emission from source s, $\mu_s$ is the total emission from source s, and $U_{total}$ is the standard deviation of total emission. The uncertainty is further expressed as coefficient of variation (standard deviation divided by corresponding mean estimation).

**Scenarios of future GHG emission from China's power system**

The baseline scenario (BS) reflects a continuation of the 2010 – 2019 growth rate in power demand and source-specific power generation. China's inter-annual monthly source-specific power generation show near-linear steady growth for the past decade (observed from activity data, SI Fig5). Based on the assumption that this near-linear steady growth is preserved for 2020 (i.e. without the disruption of COVID-19) and for the near future (till 2030). We established linear regression model as following equation (Eq.14), to generate the power generation for the BAU scenario.

$$G_{(S,M)} = \alpha_{(S,M)} Y + \beta_{(S,M)} \qquad (14)$$

$G_{(S,M)}$ stands for the power generation for month m and from energy source S. Year is the independent variable. We established the above-mentioned linear relationship for each energy source and compiled the model parameters $\alpha_{(S,M)}$ and $\beta_{(S,M)}$, with monthly power generation data from January 2010 to December 2019. We extrapolated the modelled relationship to 2020 and to 2030, under the assumption that the monthly power demand and generation continues the near-linear growth. Prediction uncertainty is indicated by the 95% prediction interval. Annual power generation is compiled as the sum of monthly source-specific power generation. Uncertainty associated with annual power generation is compiled as the propagated error from each monthly generation, expressed and calculated as the variance ($\sigma_Y$), with the following equation (Eq.15):

$$\sigma_Y = \sum \sigma_{(S,M,Y)} \qquad (15)$$

Where $\sigma_{(S,M)}$ is the variance associated with the predicted power generation by source S in month M of year Y. Established linear regression curves are visualized in SI Fig3.

The Improved energy structure (IES) assumes the same growth rate of total power demand as in the BS scenario. But for the non-coal power generation, it assumes that the growth from 2019 to 2020 will preserve for the next decade. In the IES scenario, the power demand which is not met by the increasing non-coal power, will be met by coal-fired power. The reduced demand with decarbonization (RD) scenario assumes a continuation of power demand and generation change of 2019-2020 towards 2030. This scenario assumes that the power demand and source-specific power generation will increase at the same quantity as 2019 – 2020.

*GHG emissions*. With the above-mentioned source-specific power generation and the emission factors in Table 3, we calculated source-specific GHG emissions (Eq.1). Total annual (or monthly) GHG emissions are sums of these source-specific GHG emissions. The corresponding carbon intensities are calculated with Eq.2.

# Author contributions

ZL and BZ designed the research. BZ, RG, ZD, WZ and XD performed the analyses. BZ and ZL led the writing with input from all co-authors. All co-authors reviewed and commented on the manuscript.

# Acknowledgements

Authors acknowledge comments by reviewers that helped to improve the paper.

**Data Availability Statement**

All data generated or analyzed during this study are included in this article (and its Data descriptor paper and supplementary information files).

**Code Availability Statement**

The codes generated during the current study are available from the corresponding author on reasonable request.

**Declaration**

Authors declare no competing interests.